\documentclass[journal]{IEEEtran}
\usepackage{amsmath,amssymb,amsfonts}
\usepackage{algorithmic}
\usepackage{graphicx}
\usepackage{textcomp}
\usepackage{xcolor}
\usepackage{stfloats}
\usepackage{booktabs}
\usepackage{hyperref}
\hypersetup{hypertex=true,
            colorlinks=true,
            linkcolor=blue,
            anchorcolor=blue,
            citecolor=blue}

\newtheorem{theorem}{Theorem}[section]

\newtheorem{lemma}{Lemma}
\newtheorem{remark}{Remark}

\begin{document}
\title{\LARGE \textbf{Data-Driven Robust Control for Discrete Linear Time-Invariant Systems: A Descriptor System Approach}}

\author{Jiabao He, Xuan Zhang, Feng Xu*, Junbo Tan  and Xueqian Wang
% <-this % stops a space
\thanks{J. He, F. Xu, J. Tan and X. Wang are with the Center for Intelligent Control and Telescience, Tsinghua Shenzhen International Graduate School, Tsinghua University, Shenzhen 518055, P.R.China (email: hjb18@tsinghua.org.cn; xu.feng@sz.tsinghua.edu.cn; tjblql@sz.tsinghua.edu.cn; wang.xq@sz.tsing-hua.edu.cn).}
\thanks{X. Zhang is with the Tsinghua-Berkeley Shenzhen Institute, Tsinghua Shenzhen International Graduate School, Tsinghua University, Shenzhen 518055, P.R.China, (email: xuanzhang@sz.tsing-hua.edu.cn).}
\thanks{Corresponding author: Feng Xu.}
\thanks{This work has been submitted to the IEEE for possible publication. Copyright may be transferred without notice, after which this version may no longer be accessible.}
}

%\markboth{Journal of IEEE, December~2021}%

\maketitle

\begin{abstract}
Given the recent surge of interest in data-driven control, this paper proposes a two-step method to study robust data-driven control for a parameter-unknown linear time-invariant (LTI) system that is affected by energy-bounded noises. First, two data experiments are designed and corresponding data are collected, then the investigated system is equivalently written into a data-based descriptor system with structured parametric uncertainties. Second, combined with model-based control theory for descriptor systems, state feedback controllers are designed for such data-based descriptor system, which stabilize the original LTI system and guarantee the ${H_\infty}$ performance. Finally, a simulation example is provided to illustrate the effectiveness and merits of our method.
\end{abstract}


\IEEEpeerreviewmaketitle

\section{Introduction} \label{sec1}

Recently, based on Willems et al.’s fundamental lemma in behavioral theory \cite{willems2005note}, a novel direct data-driven control (DDC) method for LTI systems was proposed in \cite{de2019formulas}, which parameterized systems by persistently exciting input and collecting state or output signals. In this way, the identification of systems' parameters is not needed, and various controllers, such as state feedback controllers and linear quadratic regulators (LQRs) can be directly designed based on those data sets. Besides, researchers in \cite{van2020data} introduced the concept of data-informativity, which answered the question that what sufficient conditions should those data sets possess for systems' analysis and control. Subsequent works have extended DDC methods in the behavioral theory to various scenarios, such as multiple data sets \cite{van2020Willems}, data-based controllability \cite{mishra2020data} and observability  \cite{eising2020data} tests, linear time-varying systems \cite{nortmann2021direct}, switched linear systems \cite{rotulo2021online}, descriptor systems \cite{schmitz2022willems}, nonlinear systems \cite{strasser2021data} and linear delay systems \cite{rueda2021data}. Moreover, data-driven model predictive control was considered in \cite{berberich2020data} and \cite{coulson2021distributionally}. More details and recent developments about DDC and the behavioral theory can be found in \cite{markovsky2021behavioral} and \cite{van2020fromdata}.

In practice, noises will inevitably appear in systems, which may deteriorate performances of controllers and even lead systems to failures. Thus, robust control is an important branch of control theories, and such importance has been highly emphasized by researchers in DDC control. Unknown measurement noises were considered in \cite{de2019formulas}, and noise levels were explicitly quantified to discuss when a stabilizing controller existed. Researchers in \cite{berberich2020robust} presented an uncertain closed-loop parameterization using noisy data, and then developed robust and ${H_\infty}$ controllers. However, those methods only provide sufficient conditions, and some of their decision variables to be computed depend on the time horizon of experiments, which means that they are slightly conservative and not applicable to large data sets. In order to solve these problems, two non-conservative methods were introduced in \cite{van2020noisy} and \cite{bisoffi2021data}. To be specific, researchers in \cite{van2020noisy} presented a matrix version of S-Lemma \cite{yakubovich1977s}, and showed how this lemma could be used to design robust, ${H_2}$ and ${H_\infty}$ controllers for LTI systems with bounded system noises. As a contrast, researchers in \cite{bisoffi2021data} addressed the robust DDC problem via Petersen’s Lemma \cite{petersen1987stabilization}, a popular tool in robust control. Besides, LQR problems using noisy data were investigated in \cite{persis2021low} and \cite{dorfler2021certainty}.

Based on the above literatures, one may conclude that the robust DDC problem for LTI systems has been well and systematically studied under the framework of behavioral theory. In this paper, we revisit this problem from a different perspective, a descriptor system approach. The core idea is to rewrite a normal state-space LTI system into a data-based descriptor system based on well-designed data experiments. Then with the help of model-based theories for descriptor systems, we show how to design robust controllers for the original system. Besides, all control conditions in this paper are cast into linear matrix inequalities (LMIs).

We consider that our method is attractive for the following reasons: (\uppercase\expandafter{\romannumeral1}) Unlike other related research, the details of data experiments are introduced in this paper, including how to design input signals and collect corresponding state and output sequences, which means that our method is easier to implement. (\uppercase\expandafter{\romannumeral2}) Comparing with \cite{berberich2020robust} and \cite{van2020noisy} which assume that the system's output matrices should be available when designing ${H_\infty}$ controllers, all system matrices in our paper are unknown, which means that our method is more general. (\uppercase\expandafter{\romannumeral3}) All decision variables in this paper are independent with the time horizon of experiments, which means that our method is more tractable and flexible to large data sets and large-scale systems. (\uppercase\expandafter{\romannumeral4}) As far as we know, all existing robust DDC methods cannot be applied to descriptor systems. Since our method is on the perspective of descriptor systems, they can be naturally extended to study the robust DDC problem for descriptor systems \cite{he2021data}.

The remainder of this paper is as follows: Section \ref{sec2} introduces preliminaries about the investigated system, and then designs two experiments to generate data. Furthermore, by replacing the system's unknown matrices with well-collected data, the original system is reformulated into a descriptor system with structural parametric uncertainties in all matrices. Section \ref{sec3} proposes robust and ${H_\infty }$ controllers' design for the data-based descriptor system. A simulation example is provided in Section \ref{sec4} to show the effectiveness of our method. Finally, conclusions are presented in Section \ref{sec5}.

\emph{Notations:} $\mathbb{R}$ and $\mathbb{C}$ are the fields of real and complex numbers. For a vector $x$, $\|x\|_2$ means the 2-norm of $x$. For a matrix $X$, the notations $\texttt{det}(X)$, $X^\mathrm{T}$, $X^{-1}$, $\|X\|_\infty$, $X \succ (\succeq) 0$ and $X \prec (\preceq) 0$ mean the determinant, the transpose, the inverse, the infinite norm, the positive (semi-positive) and negative (semi-negative) definiteness of $X$, respectively. The block matrix $\begin{bmatrix}X_1 & X_2 \\\ast & X_4 \\\end{bmatrix}$ means the symmetrical matrix $\begin{bmatrix} X_1 & X_2 \\X_2^\mathrm{T} & X_4 \\ \end{bmatrix}$. Besides, $I_{m}$, $0_m$ and $0_{m\times n}$ are identity and zero matrices of appropriate dimensions.

\section{Data Experiments and Model Representation } \label{sec2}

\subsection{Preliminaries} \label{sec2.1}
Consider the following discrete LTI system
\begin{equation} \label{E1}
\begin{split}
{x_{k + 1}} &= A{x_k} + B{u_k} + B_w{w_k},\\
{y_k} &= C{x_k} + D{u_k}
\end{split}
\end{equation}
where $x_{k}\in \mathbb{R}^{n}$ is the measurable system state, $u_{k}\in \mathbb{R}^{m}$ is the control input, $y_{k}\in \mathbb{R}^{p}$ is the system output and $w_{k}\in \mathbb{R}^{q}$ is the unmeasurable disturbance that is assumed to be energy-bounded, i.e., $\left\| {{w_k}} \right\|_2^2 \le \delta$. Besides, the matrices $A \in \mathbb{R}^{n\times n}$, $B\in \mathbb{R}^{n\times m}$ , $B_w\in \mathbb{R}^{n\times q}$, $C\in \mathbb{R}^{p\times n}$ and $D\in \mathbb{R}^{p\times m}$are unknown parameters. This paper aims to design a state feedback controller
\begin{equation} \label{E1b}
{u_k} = F{x_k}
\end{equation} based on data from well-designed experiments, such that the closed-loop system
\begin{equation} \label{E1c}
\begin{split}
{x_{k + 1}} &= (A + BF){x_k} + B_w{w_k},\\
{y_k} &= (C + DF){x_k}
\end{split}
\end{equation}
is robustly stable and satisfies the ${H_\infty }$ requirement, i.e., for a prescribed attenuation level $\gamma > 0$, the following condition holds:
\begin{equation}\label{E1d}
\sum\limits_{k = 0}^\infty  \|{y_k}\|_2^2  \le {\gamma ^2}\sum\limits_{k = 0}^\infty \|{w_k}\|_2^2.
\end{equation}

One of the core problems for DDC is how to construct a system's parameters with data. Due to the system noises ${w_k}$, it is difficult to directly identify or replace the matrices in \eqref{E1} with data, so we introduce the following descriptor system which is equivalent to the original system \eqref{E1}:
\begin{equation} \label{E2}
\begin{split}
({{s_0}I - A})^{-1}{x_{k + 1}} = &({{s_0}I - A})^{-1}A{x_k} + ({{s_0}I - A})^{-1}B{u_k} + \\
&({{s_0}I - A})^{-1}B_w{w_k},\\
{y_k} = &C{x_k} + D{u_k}
\end{split}
\end{equation}
where ${s_0} \in \mathbb{C}$ is a complex number such that $({s_0}I - A)$ is a non-singular matrix. In the following subsections, we will show how to design data experiments, such that parameters of the descriptor system \eqref{E2} can be replaced with data. Obviously, if we design a state feedback controller \eqref{E1b} based on data to stabilize the system \eqref{E2}, then the system \eqref{E1} is also  stable under the same controller, which is the core idea of this paper.

\subsection{Data Experiments} \label{sec2.2}

\emph{\textbf{Experiment 1}}: Exerting $n$ groups of sub-experiments on the nominal system \eqref{E1}. For each sub-experiment, there are $l (l\geq 1)$ steps of input sequences which satisfy
\begin{equation} \label{E3}
\sum\limits_{k = 0}^{l - 1} {{u_k}(i)}  = 0, i = 1,2,...,n.
\end{equation}
Then for the $i\texttt{-th}$ sub-experiment, we have
\begin{equation} \label{E4}
\left\{
{\begin{array}{*{20}{l}}
{{x_1}(i) = A{x_0}(i) + B{u_0}(i) + B_w{w_0}(i)}\\
{{x_2}(i) = A{x_1}(i) + B{u_1}(i) + B_w{w_1}(i)}\\
\qquad \qquad \vdots \\
{{x_l}(i) = A{x_{l - 1}}(i) + B{u_{l - 1}}(i) + B_w{w_{l - 1}}(i)}
\end{array}}
\right.
\end{equation}
and
\begin{equation} \label{E4a}
\left\{ {\begin{array}{*{20}{l}}
{{y_0}(i) = C{x_0}(i) + D{u_0}(i)}\\
{{y_1}(i) = C{x_1}(i) + D{u_1}(i)}\\
\qquad \qquad \vdots \\
{{y_{l - 1}}(i) = C{x_{l - 1}}(i) + D{u_{l - 1}}(i)}.
\end{array}} \right.
\end{equation}

First, we will show how to construct $({{s_0}I - A})^{-1}$ and $({{s_0}I - A})^{-1}A$ in \eqref{E2} with those data. Based on \eqref{E3}, after summation on both sides of \eqref{E4}, we have
\begin{equation} \label{E5}
\sum\limits_{k = 1}^l {{x_k}(i)}  = \sum\limits_{k = 0}^{l - 1} {A{x_k}(i)} + \sum\limits_{k = 0}^{l - 1} {B_w{w_k}(i)}.
\end{equation}
Since there are infinitely many ${s} \in \mathbb{C}$ such that $\texttt{det} (s I -A)\neq 0$, with an arbitrarily given number $s_0$, $(s_0 I - A)$ will be a non-singular matrix in most cases, then the equation \eqref{E5} can be rewritten as
\begin{equation} \label{E6}
\begin{split}
&( {{s_0} - 1})\sum\limits_{k = 0}^{l - 1} {{x_k}(i)}  + {x_0}(i) - {x_l}(i) = \\
&( {{s_0}I - A})\sum\limits_{k = 0}^{l - 1} {{x_k}(i)}  - \sum\limits_{k = 0}^{l - 1} {B_w{w_k}(i)}.
\end{split}
\end{equation}
Denoting vectors $n_i = {( {s_0  - 1})\sum\limits_{k = 0}^{l - 1} {{x_k}(i)}  + {x_0}(i) - {x_l}(i)}$ and $m_i = \sum\limits_{k = 0}^{l - 1} {{x_k}(i)}$, then for all $n$ groups of sub-experiments, we have
\begin{equation} \label{E7}
N = \left( {s_0 I - A} \right)M - B_wW
\end{equation}
where $N = [ {{n_1}}\ {{n_2}}\ \cdots \ {{n_n}}]$, $M = [{{m_1}}\ {{m_2}}\ \cdots\ {{m_n}}]$ and $W = [ {\sum\limits_{k = 0}^{l - 1} {{w_k}(1)} }\ {\sum\limits_{k = 0}^{l - 1} {{w_k}(2)} }\ \cdots \ {\sum\limits_{k = 0}^{l - 1} {{w_k}(n)} }]$. Since the matrix $N$ can always be observed and assigned to be non-singular, plus $\texttt{det} (s_0 I - A)\neq 0$, we have
\begin{equation} \label{E8}
{( {s_0I - A})^{ - 1}} = M{N^{ - 1}} - {( {s_0I - A})^{ - 1}}B_wW{N^{ - 1}}.
\end{equation}

Besides, the equation \eqref{E5} can also be rewritten as
\begin{equation} \label{E9}
\begin{split}
&({s_0}I - A)\sum\limits_{k = 1}^l {{x_k}(i)} = {s_0}B_w\sum\limits_{k = 0}^{l - 1} {{w_k}(i)} + \\
&A(({s_0} - 1)\sum\limits_{k = 1}^l {{x_k}(i)}  + {s_0}({x_0}(i) - {x_l}(i))).
\end{split}
\end{equation}
Similarly, by defining vectors ${v_i} = \sum\limits_{k = 1}^l {{x_k}(i)}$ and ${t_i} = ({s_0} - 1)\sum\limits_{k = 1}^l {{x_k}(i)}  + {s_0}({x_0}(i) - {x_l}(i))$, then for all $n$ groups of sub-experiments, we have
\begin{equation} \label{E10}
( {{s_0}I - A})V = AT + {s_0}B_wW
\end{equation}
where $V = [ {{v_1}}\ {{v_2}}\ \cdots \ {{v_n}}]$ and $T = [{{t_1}}\ {{t_2}}\ \cdots\ {{t_n}}]$. Like the matrix $N$, the matrix $T$ can always be assigned to be non-singular, so \eqref{E10} can be rewritten as
\begin{equation} \label{E11}
{( {{s_0}I - A})^{ - 1}}A = V{T^{ - 1}} - {s_0}{( {{s_0}I - A})^{-1}}B_wW{T^{-1}}.
\end{equation}

Second, we show how to construct the matrix $C$ in \eqref{E2}. Based on \eqref{E4a}, we have $\sum\limits_{k = 0}^{l - 1} {{y_k}(i)}  = C\sum\limits_{k = 0}^{l - 1} {{x_k}(i)}$. Furthermore, for $n$ groups of sub-experiments, it comes to
\begin{equation} \label{E11a}
Y = CX
\end{equation}
where $Y = [ {\sum\limits_{k = 0}^{l - 1} {{y_k}(1)} }\ {\sum\limits_{k = 0}^{l - 1} {{y_k}(2)} }\ \cdots \ {\sum\limits_{k = 0}^{l - 1} {{y_k}(n)} }]$ and $X = [ {\sum\limits_{k = 0}^{l - 1} {{x_k}(1)} }\ {\sum\limits_{k = 0}^{l - 1} {{x_k}(2)} }\ \cdots \ {\sum\limits_{k = 0}^{l - 1} {{x_k}(n)} }]$. Once the matrix $X$ is observed to be a non-singular matrix, then the matrix $C$ can be identified as
\begin{equation} \label{E11a}
C = C_d := YX^{-1}.
\end{equation}

\begin{remark} \label{Rem1}
In this experiment, the matrices $N$, $T$ and $X$ are required to be non-singular. Note that all those matrices contain the information of initial state $x_0(i)$ in each sub-experiment, so we can choose different initial state conditions or introduce different input signals in each sub-experiment, such as taking random signals to guarantee their invertibility.
\end{remark}

As shown in \eqref{E8}, \eqref{E11} and \eqref{E11a}, the matrices ${( {{s_0}I - A})^{ - 1}} $, ${( {{s_0}I - A})^{ - 1}}A$ and $C$ are constructed from Experiment 1, but the matrices $( {{s_0}I - A})^{ - 1}B$ and $D$ disappear in this experiment, so the following Experiment 2 shows how to construct them.

\emph{\textbf{Experiment 2}}: Exerting $m$ groups of sub-experiments on the  nominal system \eqref{E1}. In the $i\texttt{-th}$ sub-experiment, the input sequence is constant and chosen as
\begin{equation} \label{E12}
u_k^{'}(i) = {\left[{\begin{array}{*{20}{c}}{0,...,0,}&{\mathop{1}\limits_{i\texttt{-th}},}&{0,...,0}\end{array}} \right]^\mathrm{T}}, i = 1,2,...,m.
\end{equation}
Then for $l (l\geq 1)$ steps of observations in each sub-experiment, we have $x_{l}^{'}(i) = Ax_{l-1}^{'}(i) + Bu_{l-1}^{'}(i) + B_w{w_{l - 1}^{'}}(i)$ and $y_{l-1}^{'}(i) = Cx_{l-1}^{'}(i) + Du_{l-1}^{'}(i)$. After lining up outcomes of the $m$ groups, we obtain
\begin{equation} \label{E13}
{R_1} = A{R_0} + B + B_w{W_0}
\end{equation}
and
\begin{equation} \label{E13a}
Y{'} = CX{'} + D
\end{equation}
where
\begin{equation} \label{E13b}
\begin{split}
{R_1} &= \begin{bmatrix}{x_{l}^{'}(1)} & {x_{l}^{'}(2)} & \cdots & {x_{l}^{'}(m)} \\ \end{bmatrix}, \\
{R_0} &= \begin{bmatrix}{x_{l-1}^{'}(1)} & {x_{l-1}^{'}(2)} & \cdots & {x_{l-1}^{'}(m)} \\ \end{bmatrix}, \\
{W_0} &= \begin{bmatrix}{{w_{l-1}^{'}}(1)} & {{w_{l-1}^{'}}(2)} & \cdots & {{w_{l-1}^{'}}(m)} \\\end{bmatrix}, \\
Y{'} &= \begin{bmatrix}{y_{l-1}^{'}(1)} & {y_{l-1}^{'}(2)} & \cdots & {y_{l-1}^{'}(m)} \\ \end{bmatrix},\\
X{'} &= \begin{bmatrix}{x_{l-1}^{'}(1)} & {x_{l-1}^{'}(2)} & \cdots & {x_{l-1}^{'}(m)} \\ \end{bmatrix}.
\end{split}
\nonumber
\end{equation}
Furthermore, choosing the same $s_0$ as in Experiment 1 and multiplying $(s_0 I - A)^{-1}$ to both sides of \eqref{E13}, we have
\begin{equation} \label{E14}
\begin{split}
&{( {{s_0}I - A})^{ - 1}}{R_1} = {( {{s_0}I - A})^{ - 1}}A{R_0} + \\
&{( {{s_0}I - A})^{ - 1}}B + {( {{s_0}I - A})^{ - 1}}B_w{W_0}.
\end{split}
\end{equation}
Then substituting \eqref{E8} and \eqref{E11} into \eqref{E14}, we obtain
\begin{equation} \label{E15}
\begin{split}
&{( {{s_0}I - A})^{ - 1}}B = M{N^{ - 1}}{R_1} - V{T^{ - 1}}{R_0} - \\
&{( {{s_0} I - A})^{ - 1}}B_w(W({N^{ - 1}}{R_1} - s_0{T^{ - 1}}{R_0}) +W_0) .
\end{split}
\end{equation}

Besides, based on \eqref{E11a} and \eqref{E13a}, we obtain
\begin{equation} \label{E15a}
D = D_d := Y{'} - YX^{-1}X{'}.
\end{equation}
In this way, ${( {{s_0}I - A})^{ - 1}}B$ and $D$ are also constructed. However, it should be mentioned that comparing with $C$ and $D$ which are completely recovered from measurable data, there are unknown terms ${( {{s_0} I - A})^{ - 1}}B_w$, $W$ and $W_0$ in ${( {{s_0}I - A})^{ - 1}}$, ${( {{s_0}I - A})^{ - 1}}A$ and ${( {{s_0}I - A})^{ - 1}}B$, which means that these matrices can only be partially recovered, and we will show how to deal with these unknown terms in the remaining parts.

\subsection{Data-Based Model Representation} \label{sec2.3}

Since the disturbance ${w_k}$ is energy-bounded, i.e., $\left\| {{w_k}} \right\|_2^2 \le \delta$, we have ${w_k}{w_k}^\mathrm{T} \preceq \delta I$, which means that
\begin{equation} \label{E16a}
W{W^\mathrm{T}} = \sum\limits_{i = 1}^n {( {\sum\limits_{k = 0}^{l - 1} {{w_k}(i)\sum\limits_{k = 0}^{l - 1} {w_k^\mathrm{T}(i)} } })}  \preceq \delta n{l^2}I_q
\end{equation}
and
\begin{equation} \label{E16b}
W_0{W_0^\mathrm{T}} = \sum\limits_{i = 1}^m {{w_{l-1}^{'}(i)}{w_{l-1}^\mathrm{{'}T}(i)}}  \preceq \delta mI_q.
\end{equation}
In this way, the disturbance matrix $W$ and $W_0$ can be rewritten as
\begin{equation} \label{E16c}
W = l\sqrt {\delta n} \Delta _F, W_0 = \sqrt {\delta m} \Delta _F \begin{bmatrix}{I_m} \\0_{(n-m)\times m}  \\\end{bmatrix}
\end{equation}
where the uncertainty $\Delta _{F}\in \mathbb{R}^{q\times n}$ satisfies ${\Delta _F}\Delta _F^\mathrm{T} \preceq I_q$. Furthermore, combining \eqref{E8}, \eqref{E11}, \eqref{E11a}, \eqref{E15} and \eqref{E15a} with \eqref{E16c}, the model-based descriptor system \eqref{E2} can be replaced with the following data-based descriptor system
\begin{equation} \label{E17}
\begin{split}
( {{E_d} + \Delta {E_d}}){x_{k + 1}} = &( {{A_d} + \Delta {A_d}}){x_k} + \\
&( {{B_d} + \Delta {B_d}} ){u_k}+ B_{wd}{w_k}, \\
{y_k} = &C_d{x_k} + D_d{u_k}
\end{split}
\end{equation}
where the nominal matrices ${E_d} = M{N^{ - 1}}$, ${A_d} = V{T^{ - 1}}$, ${B_d} = M{N^{ - 1}}{R_1} - V{T^{ - 1}}{R_0}$, $C_d = YX^{-1}$ and $D_d = Y{'} - YX^{-1}X{'}$; the structured parametric uncertainties $\Delta {E_d} = B_{wd}{\Delta _F}{K_e}$, $\Delta {A_d} = B_{wd}{\Delta _F}{K_a}$ and $\Delta B_d = B_{wd}{\Delta _F}{K_b}$, with $K_e =  - l\sqrt {\delta n}{N^{ - 1}}$, $K_a =  - s_0l\sqrt {\delta n}{T^{ - 1}}$, ${K_b} =  - l\sqrt {\delta n}( {{N^{ - 1}}{R_1} - s_0{T^{ - 1}}{R_0}) - \sqrt {\delta m}\begin{bmatrix}{I_m} \\0_{(n-m)\times m} \\\end{bmatrix}}$ and $B_{wd} = {( {{s_0}I - A})^{ - 1}}B_w$.

Furthermore, since $( {{s_0}I - A})^{ - 1} = ( {{E_d} + \Delta {E_d}})$ is non-singular, by defining the augmented state vector ${{\hat x}_k} = \begin{bmatrix}{{x_k^\mathrm{T}}}&{{x_{k + 1}^\mathrm{T}}} \\\end{bmatrix}^\mathrm{T}$, the descriptor system \eqref{E17} is equivalent to the following augmented descriptor system
\begin{equation} \label{E18}
\begin{split}
\hat E_d{{\hat x}_{k + 1}} &= (\hat A_d + \Delta \hat A_d){{\hat x}_k} + (\hat B_d + \Delta \hat B_d){u_k} + \hat B_w{w_k}, \\
{y_k} &=\hat C_d{{\hat x}_k} + \hat D_d{u_k}
\end{split}
\end{equation}
where $\hat E_d = \begin{bmatrix}I_n&0_n\\0_n&0_n \\\end{bmatrix}$, $\hat A_d = \begin{bmatrix}0_n&I_n\\A_d&{ -  E_d}\\\end{bmatrix}$, $\hat B_d =\begin{bmatrix}0_{n\times m}\\B_d \\\end{bmatrix}$,
$\Delta \hat A_d = \begin{bmatrix}0_n&0_n\\{\Delta A_d}&{ - \Delta E_d}\\\end{bmatrix} = \hat B_w {\Delta _F}{{\hat K}_a}$, ${{\hat K}_a} = \begin{bmatrix}{{K_a}}&{ - {K_e}}\\ \end{bmatrix}$,
$\Delta \hat B_d = \begin{bmatrix}0_{n\times m}\\\Delta B_d \\\end{bmatrix} = \hat B_w {\Delta _F}{{\hat K}_b}$, ${\hat K}_b = K_b$, $\hat B_w = \begin{bmatrix}0_{n\times q}\\B_{wd} \\\end{bmatrix}$, $\hat C_d = \begin{bmatrix}{{C_d}}&0_{p\times n}\\ \end{bmatrix}$ and $\hat D_d = D_d$.

\section{Controllers Design} \label{sec3}

This section aims to design controllers for the descriptor system \eqref{E18}, so that the systems \eqref{E17}, \eqref{E18} and the original system \eqref{E1} are robustly stable and reach ${H_\infty }$ performance \eqref{E1d}. Luckily, when all parameters of the descriptor system \eqref{E18} are available, there have been fruitful model-based works to study how to design robust and ${H_\infty }$ controllers \cite{ji2007robust,mao2012robust}, which provide references for our data-based methods. As a matter of fact, most of parameters in \eqref{E18} have been replaced by available data, except for $\hat B_w$, so we will show how to amend those model-based theories to eliminate $\hat B_w$.

\subsection{Robust Controller Design} \label{sec3.1}

For the descriptor system \eqref{E18}, we have the following model-based robust control theory:
\begin{lemma} [\cite{mao2012robust}] \label{Lem1}
The descriptor system \eqref{E18} is robustly stable under the state feedback controller \eqref{E1b}, if and only if there exists a positive definite matrix $P\in \mathbb{R}^{{2n}\times {2n}}$, matrices $Q\in \mathbb{R}^{{2n}\times {n}}$, $Z\in \mathbb{R}^{{m}\times {n}}$, $H = \begin{bmatrix}K&0_n\\{{H_3}}&{{H_4}}\\\end{bmatrix}$, $G = \begin{bmatrix}K&0_n\\{{G_3}}&{{G_4}}\\\end{bmatrix}$, $K, H_3, H_4, G_3, G_4 \in \mathbb{R}^{{n}\times {n}}$ and a positive scalar $\varepsilon > 0$, such that the following LMI condition holds:
\begin{equation} \label{E20}
\Phi  = \begin{bmatrix}
          {\Phi _{11}} & {\Phi _{12}} & ({{{\hat K}_a}H + {{\hat K}_b}Z\hat I})^\mathrm{T} \\
          * & { - G - {G^\mathrm{T}} + P} & ({{{\hat K}_a}G + {{\hat K}_b}Z\hat I})^\mathrm{T} \\
          * & * & { - \varepsilon I_n} \\
        \end{bmatrix}\prec 0
\end{equation}
where $\hat I = \begin{bmatrix}I_n&0_n \\\end{bmatrix}$, $S = \begin{bmatrix}0_n&I_n \\\end{bmatrix}^\mathrm{T}$,
\begin{equation} \label{E20a}
\begin{split}
{\Phi _{11}} = &( {\hat A_d - \hat E_d})H + {H^\mathrm{T}}{{( {\hat A_d - \hat E_d})}^\mathrm{T}} + \\
&\hat B_d Z\hat I + {{( {\hat B_dZ\hat I})}^\mathrm{T}} + \varepsilon \hat B_w{{\hat B_w}^\mathrm{T}}, \\
{\Phi _{12}} = &{{\hat E}_d}P + Q{S^{\rm{T}}} - {H^{\rm{T}}} + ({{\hat A}_d} - {{\hat E}_d})G + {{\hat B}_d}Z\hat I,
\end{split}
\nonumber
\end{equation}
and the feedback gain is given as $F = Z{K^{ - 1}}$.
\end{lemma}

Note that in \eqref{E20}, all system's parameters have been replaced by data, except for $B_{wd} = {\left( {{s_0}E - A} \right)^{ - 1}}B_w$, so Lemma \ref{Lem1} cannot be directly used. However, since $B_{wd}{ B_{wd}^\mathrm{T}} \succeq 0$, we obtain the following Theorem \ref{Th1} which provides a sufficient condition relative to Lemma \ref{Lem1}:
\begin{theorem} \label{Th1}
The descriptor system \eqref{E18} is robustly stable under the state feedback controller \eqref{E1b}, if there exists a positive definite matrix $P\in \mathbb{R}^{{2n}\times {2n}}$, matrices $Q\in \mathbb{R}^{{2n}\times {n}}$, $Z\in \mathbb{R}^{{m}\times {n}}$, $H = \begin{bmatrix}K&0_n\\{{H_3}}&{{H_4}}\\\end{bmatrix}$, $G = \begin{bmatrix}K&0_n\\{{G_3}}&{{G_4}}\\\end{bmatrix}$, $K, H_3, H_4, G_3, G_4 \in \mathbb{R}^{{n}\times {n}}$ and a positive scalar $\varepsilon > 0$, such that the following LMI condition holds:
\begin{equation} \label{E21}
\Phi^{'}  = \begin{bmatrix}
          {\Phi _{11}^{'}} & {\Phi _{12}} & ({{{\hat K}_a}H + {{\hat K}_b}Z\hat I})^\mathrm{T} \\
          * & { - G - {G^\mathrm{T}} + P} & ({{{\hat K}_a}G + {{\hat K}_b}Z\hat I})^\mathrm{T} \\
          * & * & { - \varepsilon I_n} \\
        \end{bmatrix}\prec 0
\end{equation}
where ${\Phi _{11}^{'}} = ( {\hat A_d - \hat E_d})H + {H^\mathrm{T}}{{( {\hat A_d - \hat E_d})}^\mathrm{T}} + \hat B_d Z\hat I + {{( {\hat B_dZ\hat I})}^\mathrm{T}}$ and the feedback gain is given as $F = Z{K^{ - 1}}$.
\end{theorem}

\emph{Proof:} Based on Lemma \ref{Lem1} and $\hat B_w{{\hat B_w}^\mathrm{T}} \succeq 0$, it is straightforward to obtain the results in Theorem \ref{Th1}. \hfill $\square$

\subsection{${H_\infty }$ Controller Design} \label{sec3.2}

For the descriptor system \eqref{E18}, when $u_k = 0$ and $\Delta _F = 0$, we have the following model-based ${H_\infty }$ control theory:
\begin{lemma} [\cite{ji2007robust}] \label{Lem2}
For a prescribed scalar $\gamma >0$, the system \eqref{E18} with $u_k = 0$ and $\Delta _F = 0$ is stable and satisfies the ${H_\infty }$ condition \eqref{E1d}, if and only if there exists a positive definite matrix $\hat P \in {\mathbb{R}^{2n \times 2n}}$ and a matrix $\hat S \in {\mathbb{R}^{2n \times n}}$ such that the following LMI condition holds:
\begin{equation} \label{E22}
\Theta  = \begin{bmatrix}
{{\Theta _{11}}}&{\hat S{{\hat R}^\mathrm{T}}{{\hat B}_w}}&{{{\hat A}_d}^\mathrm{T}\hat P}&{{{\hat C}_d}^\mathrm{T}}\\
*&{ - {\gamma ^2}I_q}&{{{\hat B}_w}^\mathrm{T}\hat P}&0_{q\times p}\\
*&*&{ - \hat P}&0_{2n\times p}\\
*&*&*&{ - I_{p}} \\
\end{bmatrix} \prec 0
\end{equation}
where ${\Theta _{11}} = \hat S{{\hat R}^\mathrm{T}}{{\hat A}_d} + {{\hat A}_d}^\mathrm{T}\hat R{{\hat S}^\mathrm{T}} - {\rm{ }}{{\hat E}_d}^\mathrm{T}\hat P{\rm{ }}{{\hat E}_d}$, and $\hat R \in {\mathbb{R}^{2n \times n}}$ is any matrix with full column rank and satisfies ${\hat E}_d^\mathrm{T}\hat R = 0$.
\end{lemma}

Besides, we need the following Petersen’s Lemma to deal with the uncertainty $\Delta _F$.
\begin{lemma} [\cite{petersen1987stabilization}] \label{Lem3}
Given real matrices $\hat Z$, $\hat X$, and $\hat Y$ of appropriate dimensions and with $\hat Z$ being symmetrical, then $\hat Z + \hat X\Delta _F\hat Y + \hat Y^{\mathrm{T}}\Delta _F^{\mathrm{\mathrm{T}}}\hat X^{\mathrm{T}} \prec 0$ for all $F$ satisfying $\Delta _F^{\mathrm{T}}\Delta _F \preceq I$, if and only if there exists a real scalar $\varepsilon > 0$ such that
\begin{equation} \label{E23}
\hat Z + \varepsilon \hat X{\hat X^\mathrm{T}} + \frac{1}{\varepsilon }{\hat Y^\mathrm{T}}\hat Y \prec 0.
\end{equation}
\end{lemma}

\begin{theorem} \label{Th2}
For a prescribed scalar $\gamma >0$, the system \eqref{E18} with the state feedback controller \eqref{E1b} is stable and satisfies the ${H_\infty }$ condition \eqref{E1d}, if there exists positive definite matrices $P_1, P_4 \in {\mathbb{R}^{n \times n}}$, matrices $S_1, S_2 \in {\mathbb{R}^{n \times n}}$, $K_1 \in {\mathbb{R}^{m \times n}}$ and a real scalar $\varepsilon > 0$, such that the following LMI holds:
\begin{equation} \label{E24}
\Psi  =  \begin{bmatrix}
{{\Psi _{11}}}&0_{2n\times p}&{{\Psi _{13}}}&{{\Psi _{14}}}\\
*&{ - {\gamma ^2}I_{p}}&{{\Psi _{23}}}&0_{p\times n}\\
*&*&{ - \hat P}&0_{2n\times n}\\
*&*&*&{ - \varepsilon I_{n}} \\
\end{bmatrix} \prec 0
\end{equation}
where $\Psi_{11}  = \begin{bmatrix}
{{S_1} + {S_1}^\mathrm{T} - {P_1}}&{ - {S_1}{E_d}^\mathrm{T} + {S_2}^\mathrm{T}}\\
{ - {E_d}{S_1}^\mathrm{T} + {S_2}}&{ - {E_d}{S_2}^\mathrm{T} - {S_2}{E_d}^\mathrm{T}} \\
\end{bmatrix}$,
$\Psi_{13}  = \begin{bmatrix}
0_n&{ - {P_4}}\\
{{A_d}{P_1} + {B_d}{K_1}}&{ - {E_d}{P_4}} \\
\end{bmatrix}$,
$\Psi_{14}  = \begin{bmatrix}
{ - {S_1}{K_e}^\mathrm{T}}\\
{ - {S_2}{K_e}^\mathrm{T}} \\
\end{bmatrix}$,
$\Psi_{23}  = \begin{bmatrix}
{{C_d}{P_1} + {D_d}{K_1}}&0_{p\times n}\\
\end{bmatrix}$ and the feedback gain $F = {K_1}{P_1}^{ - 1}$.
\end{theorem}

\emph{Proof:} First, we consider the closed-loop system \eqref{E18} with the state-feedback controller ${u_k} = F{x_k} = \begin{bmatrix}
F&0_{m\times n}\\\end{bmatrix}{\hat x_k}$ and the uncertainty $\Delta _F = 0$. In this way, we have ${{\hat A}_{dc}} = \begin{bmatrix}
0_n&I_n\\
{{A_d} + {B_d}F}&{ - {E_d}} \\
\end{bmatrix}$ and ${{\hat C}_{dc}} = \begin{bmatrix}
{{C_d} + {D_d}F}&0_{p\times n}\\\end{bmatrix}$. After replacing ${{\hat A}_{d}}$ and ${{\hat C}_{d}}$ with ${{\hat A}_{dc}}$ and ${{\hat C}_{dc}}$ in \eqref{E22}, we conclude that the closed-loop system \eqref{E18} with $\Delta _F = 0$ is stable with disturbance attenuation $\gamma$, if and only if there exists a positive definite matrix $\hat P \in {\mathbb{R}^{2n \times 2n}}$ and a matrix $\hat S \in {\mathbb{R}^{2n \times n}}$, such that the following LMI holds:
\begin{equation} \label{E25}
\Xi  =  \begin{bmatrix}
{{\Xi _{11}}}&{\hat S{{\hat R}^\mathrm{T}}{{\hat B}_w}}&{{{\hat A}_{dc}}^\mathrm{T}\hat P}&{{{\hat C}_{dc}}^\mathrm{T}}\\
*&{ - {\gamma ^2}I_q}&{{{\hat B}_w}^\mathrm{T}\hat P}&0_{q\times p}\\
*&*&{ - \hat P}&0_{2n\times p}\\
*&*&*&{ - I_{p}} \\
\end{bmatrix} \prec 0
\end{equation}
where ${\Xi _{11}} = \hat S{{\hat R}^\mathrm{T}}{{\hat A}_{dc}} + {{\hat A}_{dc}}^\mathrm{T}\hat R{{\hat S}^\mathrm{T}} - {\rm{ }}{{\hat E}_d}^\mathrm{T}\hat P{\rm{ }}{{\hat E}_d}$, and $\hat R \in {\mathbb{R}^{2n \times n}}$ is any matrix with full column rank and satisfies ${\hat E}_d^\mathrm{T}\hat R = 0$.

Second, it is well-known that the ${H_\infty }$ condition \eqref{E1d} is equivalent to ${\left\| {{H_{yw}}(s)} \right\|_\infty } \leq \gamma $ where $s \in \mathbb{C}$ and ${H_{yw}}(s) = {{\hat C}_{dc}}{(s{{\hat E}_d} - {{\hat A}_{dc}})^{ - 1}}{{\hat B}_w}$. Since ${\left\| {{\hat C}_{dc}}{(s{{\hat E}_d} - {{\hat A}_{dc}})^{ - 1}}{{\hat B}_w} \right\|_\infty } = {\left\| {{\hat B}_w}^\mathrm{T}{(s{{\hat E}_d}^\mathrm{T} - {{\hat A}_{dc}}^\mathrm{T})^{ - 1}}{{\hat C}_{dc}}^\mathrm{T} \right\|_\infty } := {\left\| {{H_{y\eta}}(s)} \right\|_\infty }$, as long as the stability and ${H_\infty }$ performance are concerned, we can consider the following system instead of \eqref{E18}:
\begin{equation} \label{E26}
\begin{split}
{{\hat E}_d}^\mathrm{T}{\zeta _{k + 1}} &= {{\hat A}_{dc}}^\mathrm{T}{\zeta _k} + {{\hat C}_{dc}}^\mathrm{T}{\eta _k}, \\
{y_k} &= {{\hat B}_w}^\mathrm{T}{\zeta _k}
\end{split}
\end{equation}
where ${\zeta _k}$ and ${\eta _k}$ are the well-defined state vector and disturbance vector, respectively. In this way, by replacing ${{\hat E}_d}$, ${{\hat A}_{dc}}$, ${{\hat B}_w}$ and ${{\hat C}_{dc}}$ in \eqref{E25} with ${{\hat E}_d}^\mathrm{T}$, ${{\hat A}_{dc}}^\mathrm{T}$, ${{\hat C}_{dc}}^\mathrm{T}$ and ${{\hat B}_w}^\mathrm{T}$, we obtain the following LMI condition which also guarantees that the closed-loop system \eqref{E18} with $\Delta _F = 0$ is stable with disturbance attenuation $\gamma$:
\begin{equation} \label{E27}
\Xi^{'} = \begin{bmatrix}
{{\Xi _{11}^{'}}}&{\hat S{{\hat R}^\mathrm{T}}{{\hat C}_{dc}}^\mathrm{T}}&{{{\hat A}_{dc}}\hat P}&{{{\hat B}_w}}\\
*&{ - {\gamma ^2}I_p}&{{{\hat C}_{dc}}\hat P}&0_{p\times q}\\
*&*&{ - \hat P}&0_{2n\times q}\\
*&*&*&{ - I_{q}} \\
\end{bmatrix} \prec 0
\end{equation}
where ${\Xi _{11}^{'}} = \hat S{{\hat R}^\mathrm{T}}{{\hat A}_{dc}}^\mathrm{T} + {{\hat A}_{dc}}\hat R{{\hat S}^\mathrm{T}} - {{\hat E}_d}\hat P{{\hat E}_d}^\mathrm{T}$. Furthermore, a sufficient condition for \eqref{E27} is that
\begin{equation} \label{E28}
\Xi^{''} = \begin{bmatrix}
{{\Xi _{11}^{'}}}&{\hat S{{\hat R}^\mathrm{T}}{{\hat C}_{dc}}^\mathrm{T}}&{{{\hat A}_{dc}}\hat P}\\
*&{ - {\gamma ^2}I_p}&{{{\hat C}_{dc}}\hat P}\\
*&*&{ - \hat P}\\
\end{bmatrix} \prec 0.
\end{equation}

Third, we deal with the uncertainty $\Delta _F \neq 0$. For the closed-loop system \eqref{E18}, the uncertain part $\Delta {\hat A_{dc}} = \begin{bmatrix}
0_n&0_n\\
{\Delta {A_d} + \Delta {B_d}F}&{ - \Delta {E_d}}\\
\end{bmatrix} = {{\hat B}_w}\Delta _F{\hat K_{ac}}$ where ${\hat K_{ac}} = \begin{bmatrix}
{{K_a} + {K_b}F}&{ - {K_e}}\\
\end{bmatrix}$. After replacing ${{\hat A}_{dc}}$ in \eqref{E28} with ${{\hat A}_{dc}} + \Delta {{\hat A}_{dc}}$, and with the help of Lemma \ref{Lem3}, we conclude that the closed-loop system \eqref{E18} is stable and satisfies the ${H_\infty }$ condition \eqref{E1d}, if there exists a positive definite matrix $\hat P \in {\mathbb{R}^{2n \times 2n}}$ and a matrix $\hat S \in {\mathbb{R}^{2n \times n}}$, such that the following LMI holds:
\begin{equation} \label{E29}
\Upsilon  =\begin{bmatrix}
{{\Upsilon _{11}}}&{\hat S{{\hat R}^\mathrm{T}}{{\hat C}_{dc}}^\mathrm{T}}&{{{\hat A}_{dc}}\hat P}&{\hat S{{\hat R}^\mathrm{T}}{{\hat K}_{ac}}^\mathrm{T}}\\
*&{ - {\gamma ^2}I_p}&{{{\hat C}_{dc}}\hat P}&0_{p\times n}\\
*&*&{ - \hat P}&0_{2n\times n}\\
*&*&*&{ - \varepsilon I_n}\\
\end{bmatrix} \prec 0
\end{equation}
where ${\Upsilon _{11}} = \hat S{{\hat R}^\mathrm{T}}{{\hat A}_{dc}}^\mathrm{T} + {{\hat A}_{dc}}\hat R{{\hat S}^\mathrm{T}} - {{\hat E}_d}\hat P{\rm{ }}{{\hat E}_d}^\mathrm{T} + \varepsilon {{\hat B}_w}{{\hat B}_w}^\mathrm{T}$. After taking $\hat P = \begin{bmatrix}
{{P_1}}&0_n\\
0_n&{{P_4}}\\
\end{bmatrix}$, $\hat S = \begin{bmatrix}
{{S_1}}\\
{{S_2}}\\
\end{bmatrix}$, $\hat R = \begin{bmatrix}
0_n\\
I_n\\
\end{bmatrix}$, $K_1 = FP_1$ and abandoning $\varepsilon {{\hat B}_w}{{\hat B}_w}^\mathrm{T}$, we can see that \eqref{E24} is a sufficient condition for \eqref{E29}, so the proof is completed.
\hfill $\square$

\section{Simulation}\label{sec4}

In this section, we will illustrate the effectiveness of our methods via a simulation example coming from \cite{van2020noisy}, i.e., considering an unstable system of the form \eqref{E1} where parameters are given as
$A =\begin{bmatrix}
0.850&-0.038&-0.380\\
0.735&0.815&1.594\\
-0.664&0.697&-0.064
\\\end{bmatrix}$, $B =\begin{bmatrix}
1.431&0.705\\
1.620&-1.129\\
0.913&0.369
\\\end{bmatrix}$, $B_w = I_3$, $C = \begin{bmatrix}
1&0&0\\
0&1&0\\\end{bmatrix}$ and $D = I_2$. For all experiments, we choose $s_0 = 0.5$ and the step length $l = 4$. As for the bound of noises, we take $\delta  = 0.2$. Inputs in the Experiment 1 and the initial system state $x_0$ are given by MATLAB function $\texttt{rand}()$. Besides, all LMI conditions are solved by YALMIP \cite{lofberg2004} in MATLAB.

(\uppercase\expandafter{\romannumeral1}) Robust controller design: After exerting Experiments 1 and 2 and collecting corresponding data, we construct the data-based system's parameters in \eqref{E17}. Then solving the LMI condition \eqref{E21}, we obtain a robust controller gain as $F_R = \begin{bmatrix}
0.3815&-0.6629&-0.5368\\
-0.1548&0.6346&1.2579\\\end{bmatrix}$. It can be checked that all eigenvalues of the matrix $(A + BF_R)$ are in the unite disk, so the system is robustly stable.

\begin{figure}
\centering
\includegraphics[scale = 0.6]{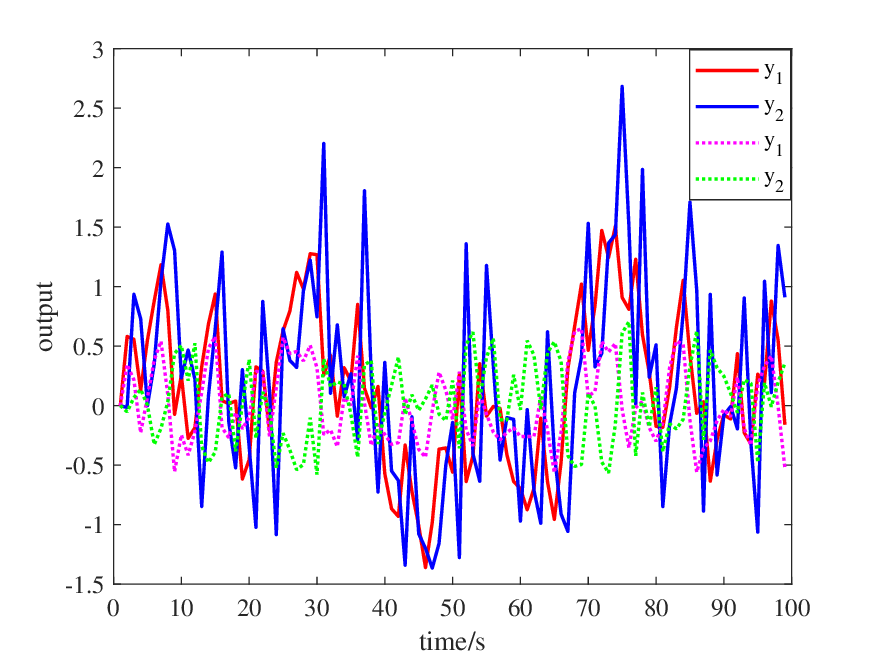}
\caption{Output sequences under the robust controller (solid line) and the ${H_\infty }$ controller (dashed line).}
\label{F1}
\end{figure}

(\uppercase\expandafter{\romannumeral2}) ${H_\infty }$ controller design: Similarly, after exerting Experiments 1 and 2 and collecting corresponding data, solving the LMI condition \eqref{E24} with the given disturbance attenuation $\gamma = 0.5$, we obtain the ${H_\infty}$ controller gain as $F_H = \begin{bmatrix}
-0.1788&-0.4381&-0.3199\\
-0.2071&0.2021&0.9403\\\end{bmatrix}$. It can also be checked that all eigenvalues of the matrix $(A + BF_H)$ are in the unite disk, so the system is stable. Besides, Figure \ref{F1} shows the system's output sequences under the robust controller $F_R$ and the ${H_\infty }$ controller $F_H$ with the same zero initial state condition, respectively. It can be seen that comparing with the robust controller, the disturbance attenuation is limited to a smaller level by the ${H_\infty }$ controller.

(\uppercase\expandafter{\romannumeral3}) Comparison with other methods: In this part, we compare our methods with the non-conservative method based on S-lemma in \cite{van2020noisy}. To illustrate, we take six different noise levels $\delta \in \{0.5, 1, 1.5, 2, 2.2, 2.4\}$. For each noise level, we generate 100 data sets by Experiments 1 and 2. Then we check the condition \eqref{E21} via these data sets. For each noise level, we record the percentage of data sets from which a stabilizing controller could be found. From Table \ref{tab1}, we can see that comparing with the non-conservative method in \cite{van2020noisy}, the percentage of our methods is smaller. The reason is that we abandon $B_{wd}{ B_{wd}^\mathrm{T}}$ in Theorem \ref{Th1} as $B_{wd} = {\left( {{s_0}E - A} \right)^{ - 1}}B_w$ is unavailable, which leads to a slightly conservative result. However, as we have emphasized in Section \ref{sec1}, comparing with \cite{van2020noisy}, the advantages of our methods are mainly in two aspects, one is about the details of data experiments, and the other is that the matrices $C$ and $D$ are not needed to be known when designing ${H_\infty }$ controllers.
\begin{table}[htbp]
\centering
\caption{Percentages of data sets for which the robust controller could be found.}
\begin{tabular}{lllllll}
    \toprule
$\delta$ &0.5&1&1.5&2&2.2&2.4\\
    \midrule
 Theorem 14 in \cite{van2020noisy} &100\%&96\%&90\%&82\%&75\%&73\%\\
 Theorem \ref{Th1}&95\%&94\%&81\%&77\%&70\%&62\%\\
   \bottomrule
  \end{tabular}
 \label{tab1}
\end{table}

\section{Conclusions}\label{sec5}

This paper proposes a novel DDC method to study the robust control for LTI systems with energy-bounded disturbances. First, we transform the original system into an equivalent system in the form of descriptor systems. Second, two groups of data experiments are designed, which allow us to replace the equivalent system's parameters with well-collected data. In this way, the original normal LTI system is reformulated into a descriptor system with structural parametric uncertainties in all matrices. Third, we turn to the model-based control theory of descriptor systems, so that the robust controller and ${H_\infty }$ controller are designed as LMIs, which permit the original system to be robustly stable and reach ${H_\infty }$ performance. In the future, we will extend methods in this paper to study robust DDC problems of descriptor systems.

\section*{Acknowledgment}\label{sec6}

This work was supported by the National Natural Science Foundation of China (No.62003186, No.62103225), the Natural Science Foundation of Guangdong, China (2021A1515012628), and the China Postdoctoral Science Foundation (No. 2020M680565).

%\section*{Acknowledgment}\label{sec6}
%\fi
\bibliographystyle{IEEEtran}
\bibliography{IEEE-J}

\end{document}